\documentclass[journal,10pt]{IEEEtran}
\usepackage[T1]{fontenc}

\usepackage[utf8]{inputenc}
\usepackage{acronym}
\usepackage{cite}
\usepackage{float}
\usepackage{multirow}
\usepackage{centernot}

\restylefloat{table}

\ifCLASSINFOpdf
   \usepackage[pdftex]{graphicx}
\else
   \usepackage[dvips]{graphicx}
\fi

\usepackage{amsmath}
\usepackage{amsfonts}
\usepackage{algorithmic}
\usepackage{algorithm}
\usepackage{color}
\usepackage{mathtools}
\DeclarePairedDelimiter{\floor}{\lfloor}{\rfloor}

\newtheorem{theorem}{Theorem}

\def\B{\boldsymbol}
\newcommand{\eul}{{\text{e}}}
\newcommand{\imj}{{\text{j}}}

\makeatletter
\newcommand*\titleheader[1]{\gdef\@titleheader{#1}}
\AtBeginDocument{%
	\let\st@red@title\@title
	\def\@title{%
		\bgroup\normalfont\large\centering\@titleheader\par\egroup
		\vskip1.5em\st@red@title}
}
\makeatother

\title{FDD Channel Estimation via Covariance Identification in Wideband Massive MIMO Systems}
\titleheader{This work has been submitted to the IEEE for possible publication. Copyright may be transferred without notice, after which this version may no longer be accessible}
\author{\IEEEauthorblockN{}
	\IEEEauthorblockA{José González-Coma$^*$, Pedro Suárez-Casal$^\star$, Paula M. Castro$^\star$, Luis Castedo$^*$, and Michael Joham$^\dagger$
	\\ University of A Coruña, $^*$CITIC,
	$^\star$Department of Computer Engineering\\
	$^\dagger$Technische Universität München, Associate Institute for Signal Processing\\
	\{jose.gcoma, pedro.scasal, paula.castro, luis.castedo\}@udc.es, joham@tum.de}
}
	
\begin{document}
\maketitle

\acrodef{ADC}[ADC]{Analog-to-Digital Converter}
\acrodef{AO}[AO]{Alternating Optimization}
\acrodef{AoA}[AoA]{Angle of Arrival}
\acrodef{AoD}[AoD]{Angle of Departure}
\acrodef{APN}[APN]{Analog Precoding Network}
\acrodef{ASK}[ASK]{Amplitude-Shift Keying}
\acrodef{AWGN}[AWGN]{Additive White Gaussian Noise}
\acrodef{BER}[BER]{Bit Error Ratio}
\acrodef{BF}[BF]{Beamforming}
\acrodef{BFN}[BFN]{Beamforming Network}
\acrodef{BS}[BS]{Base Station}
\acrodef{COMP}[COMP]{Covariance OMP}
\newacro{CSI}{Channel State Information}
\acrodef{DAC}[DAC]{Digital to Analog Converter}
\acrodef{DCS}[DCS]{Digital Communication System}
\acrodef{DCOMP}[DCOMP]{Dynamic COMP}
\newacro{DFT}{Discrete Fourier Transform}
\acrodef{DOA}[DOA]{Direction Of Arrival}
\acrodef{ESD}[ESD]{Energy Spectral Density}
\newacro{FDD}{Frequency-Division Duplex}
\acrodef{FSK}[FSK]{Frequency-Shift Keying}
\acrodef{FT}[FT]{Fourier Transform}
\acrodef{HT}[HT]{Hilbert Transform}
\acrodef{IL}[IL]{Insertion Losses}
\acrodef{ISI}[ISI]{Inter-Symbol Interference}
\acrodef{JSDM}[JSDM]{Joint Spatial Division and Multiplexing}
\acrodef{LBFN}[LBFN]{Linear Beamforming Network}
\acrodef{LLF}[LLF]{Log-Likelihood function}
\acrodef{LMD}[LMD]{Linearly Modulated Digital}
\acrodef{LOS}[LOS]{Line-of-sight}
\newacro{MAP}{Maximum A Posteriori}
\acrodef{MIMO}[MIMO]{Multiple-Input Multiple-Output}
\newacro{ML}{Maximum Likelihood}
\newacro{MMSE}{Minimum Mean Squared Error}
\acrodef{MMV}[MMV]{Multiple Measurement Vector}
\acrodef{mmWave}[mmWave]{millimeter wave}
\acrodef{MRC}[MRC]{Maximum Ratio Combining}
\newacro{MSE}{Mean Squared Error}
\acrodef{MUSIC}[MUSIC]{MUltiple SIgnal Classification}
\acrodef{NLOS}[NLOS]{Non Line-of-sight}
\acrodef{NMSE}{Normalized Mean Squared Error}
\acrodef{OFDM}[OFDM]{Orthogonal Frequency-Division Multiplexing}
\acrodef{OMP}[OMP]{Orthogonal Matching Pursuit}
\acrodef{OSMP}[OSMP]{Orthogonal Subspace Matching Pursuit}
\acrodef{PA}[PA]{Power Amplifier}
\acrodef{PS}[PS]{Phase Shifter}
\acrodef{PSK}[PSK]{Phase-Shift Keying}
\acrodef{QAM}[QAM]{Quadrature Amplitude Modulation}
\acrodef{RF}[RF]{Radio Frequency}
\acrodef{RFC}[RFC]{Rayleigh Fading Channel}
\acrodef{SDMA}[SDMA]{space-division multiple access}
\acrodef{SER}[SER]{Symbol Error Rate}
\acrodef{SLL}[SLL]{Side-Lobe Level}
\acrodef{SOMP}[SOMP]{BUSCAR}
\acrodef{SR}[SR]{Sideband Radiation}
\acrodef{SS}[SS]{Spatial Smoothing}
\acrodef{SNR}[SNR]{Signal-to-Noise Ratio}
\newacro{TDD}{Time-Division Duplex}
\acrodef{TM}[TM]{Time Modulation}
\acrodef{TMA}[TMA]{Time-Modulated Array}
\acrodef{ULA}[ULA]{Uniform Linear Array}
\acrodef{UPA}[UPA]{Uniform Planar Array}
\acrodef{VGA}[VGA]{Variable Gain Amplifier}
\acrodef{VPS}[VPS]{Variable Phase Shifter}


\begin{abstract}
	A method for channel estimation in wideband massive \ac{MIMO} systems using covariance identification is developed. The method is useful for \ac{FDD} at either sub-$6$GHz or \ac{mmWave} frequency bands and takes into account the beam squint effect caused by the large bandwidth of the signals. The method relies on the slow time variation of the channel covariance matrix and allows for the utilization of very short training sequences thanks to the exploitation of the channel structure, both in the covariance matrix identification and the channel estimation stages. As a consequence of significantly reducing the training overhead, the proposed channel estimator has a computational complexity lower than other existing approaches.

\end{abstract}

\acresetall
\section{Introduction}

Massive \ac{MIMO} and \ac{mmWave} technologies are promising candidates to satisfy the demands of future wireless communication systems. A common feature of these technologies is the deployment of large antenna arrays which allow to achieve several advantages like large array gains, or supporting many spatially multiplexed streams \cite{RuPeLaLaMaEdTu13}. In turn, large beamforming gains necessary to compensate the propagation losses at \ac{mmWave} frequencies are achieved with small sized antenna apertures \cite{AlMoGoHe14}.

The benefits of these large antenna structures rely on the accuracy of the \ac{CSI}. However, acquiring the \ac{CSI} is challenging in this scenario since the channel coherence time is shared between training and data transmission stages. When the system operates in \ac{FDD} mode and the channel is estimated in the downlink, the length of the training sequence depends on the number of antennas at the \ac{BS} \cite{HaHo03}. Moreover, it is necessary that the users feed some information back to the \ac{BS} though a link usually limited in terms of data rate. Nonetheless, many current communication systems operate in this mode \cite{BjLaMa16}.
In addition, the \ac{BS} has more power available for sending the training sequences \cite{ZhCiGr17}. 

\ac{TDD} mode is the alternative where the channel estimation is performed in the uplink leveraging on the channel reciprocity between the downlink and the uplink. The main advantage is that the training overhead is proportional to the number of users \cite{Ma06}. Another argument to employ \ac{TDD} is that it avoids the feedforward stage since only the \ac{BS} needs to acquire \ac{CSI}. Nevertheless, this information is required at the user's end when they deploy multiple antennas, which is typical when operating at \ac{mmWave} frequencies. Moreover, the large number of users enabled by massive \ac{MIMO} setups leads to the so-called pilot contamination effect \cite{JoAsMaVi11} for uplink training. 

Deploying large antenna arrays also raises concerns in terms of hardware cost and power consumption, specially at \ac{mmWave} frequencies. To alleviate these concerns, hybrid architectures with an analog preprocessing network operating in the \ac{RF} domain have been proposed to reduce the number of complete \ac{RF} chains \cite{AlMoGoHe14}. This analog stage is usually implemented using a \ac{PS} network. In terms of channel estimation, relevant drawbacks of hybrid architectures derive from the stringent limitations imposed by this hardware like the reduced dimensionality of the observed signals, the lack of flexibility of the \ac{PS}, and  the frequency-flat nature of the analog network \cite{mailloux2018phased}. These restrictions reduce the potential subspace for channel estimation \cite{BjPeBuLa18}, although this limitation can be compensated with longer training overheads. 

The channel coherence time depends on several system characteristics like the propagation environment, the mobile user speed, and the carrier frequency, among others. Hence, estimating the channel might not be practical in certain setups. To reduce the training overhead, several assumptions have been considered in the literature: 

1) \textit{Reciprocity}. Some methods for narrowband \cite{NeWiUt18} and wideband \cite{GaHuDaWa16,GoRoGoCaHe18} channel estimation exploit the channel reciprocity of \ac{TDD}. Other solutions rely on angular reciprocity for \ac{FDD} systems, e.g. \cite{LuZhCaShHuQi17,ZhWaSu17}. This approach has been also applied to measured channels with a separation between uplink and downlink of $72$ MHz \cite{ZhZhSa18}. However, the performance metric neglects inter-user interference and does not provide insights about the performance in multiuser settings.

Reciprocity relies, on the one hand, on the proper calibration of systems operating in \ac{TDD} mode. Nevertheless, uplink and downlink channels are different in \ac{FDD} mode. On the other hand, a weaker assumption is angular reciprocity. This condition holds when the carrier frequency separation between the uplink and the downlink is small \cite{Hon17, ZhZhSa18, Kha18}. However, this feature might not apply in future wireless communication systems where the large signal bandwidths do not allow for small uplink-downlink frequency separation. An example scenario is a wideband \ac{mmWave} system for which the antenna array responses are, in addition, affected by the so-called beam squint effect \cite{GaKlNaVi11}. Moreover, inaccurate angular information can potentially contaminate the channel estimate, as pointed out in \cite{ShZhAlLe16}. 

2) \textit{Sparsity.} Another family of solutions rely on the channel sparsity  in the angular domain, like \cite{NeWiUt18,GaHuDaWa16,GoRoGoCaHe18}. Since the position of the \ac{BS} is usually elevated, the few scatterers around lead to small angular spread \cite{GaDaWaCh15,ShZhAlLe16,HuHuXuYa17,ZhWaSu17}. Moreover, this effect is combined with the joint channel sparsity among the users surrounded by some common local scatterers  \cite{RaLa14,ChSuLi17}. To further reduce the overhead, some schemes consider that the support remains similar or the same during a certain period of time \cite{GaDaWaCh15,HaLeLo17}. Massive \ac{MIMO} channels are not sparse in general, thus allowing to multiplex many simultaneous users \cite{BjPeBuLa18}, e.g., as observed from measurements in environments with local scatterers around the \ac{BS} \cite{GaTuEdRu12}. Contrarily, at \ac{mmWave} frequencies, the channel consists of a few propagation paths and smaller angular spreads. This claim is supported by observations in measurement campaigns (see \cite{Sa02,SuMaRa17}).

3) \textit{Prior knowledge.} Some authors consider that the second order channel statistics are known in advance. In particular, \cite{Adh13,ZhCiGr17} propose user grouping according to some common support criterion based on the channel covariance. Next, the training is designed for each group to eventually estimate the channel. This interesting scheme is termed \ac{JSDM}. In \cite{NoZoLo16}, authors employ Kalman filtering by assuming that temporal correlation is also known. Other approaches consider that the channel statistics are known only to the users \cite{ChLoBi14}, or that partial support information is available at the \ac{BS} \cite{ShZhAlLe16}.

Prior information, in particular knowledge of the channel covariance matrix, is very helpful to reduce the training overhead. Unfortunately, this information has to be acquired too in practical settings. Remarkably, the channel covariance matrix slowly varies in time compared to the channel itself \cite{NoZoLo16}. Indeed, this time interval is several orders of magnitude larger than the channel coherence time \cite{YiGeCo14}. That is to say, the channels can be assumed to be wide-sense stationary over a certain  window of time. Experimental investigations have confirmed this fact in certain scenarios like, for example,  urban macrocell \cite{IsDoAsZe13}. Covariance identification is thereby less challenging than estimating the channel itself.

Building on this idea, previous work in the literature addressed the problem of channel covariance identification. A solution for \ac{TDD} systems was presented in \cite{NeShJoUt17}, where pilots are assigned to users depending on the properties of their covariance matrices in order to avoid pilot contamination. Authors in \cite{Hon17,Fan17,MiCaSt18,Kha18} consider covariance identification and leverage angular reciprocity to infer the downlink covariance matrix. Channel sparsity is assumed in \cite{Hon17,Fan17} to identify the covariance matrix in the uplink. Next, users are grouped to design the downlink training and estimate the channels. Authors in \cite{Kha18} employ a Toeplitz covariance estimator and solve a non-negative least squares problem in the uplink. Then, a sparsifying precoder is designed to train the downlink channel.

A different strategy is to compute the maximum likelihood estimator of the channel covariance matrix by solving a convex program \cite{HaCa17,HaCa18}.  The computational complexity of these solutions is high and an extension for wideband scenarios is missing. Frequency selective channels have been considered in \cite{PaHe17,HaBaCa18}. The model is simplified in \cite{HaBaCa18} by considering approximately frequency flat second order channel statistics.  In \cite{PaHe17}, authors address covariance identification in \ac{TDD} mode for  hybrid architectures.  The proposed algorithm is an application of \ac{OMP}, and solves the \ac{MMV} problem by allowing for a dynamic sensing matrix. In addition, this method is extended to wideband scenarios under the assumption of common subcarrier support, as done in \cite{GaDaWaCh15,GoRoGoCaHe18}. This assumption is impractical for scenarios where the signal bandwidth is as large as several GHz \cite{BjPeBuLa18} because of the beam squint effect. 

Finally, note that the consideration of hybrid digital-analog architectures poses further difficulties to wideband frequency estimation. Under the usual assumption of using \acp{PS}, the analog network is frequency flat \cite{mailloux2018phased}. This is incompatible with methods employing frequency selective training sequences \cite{GaDaWaCh15,LuZhCaShHuQi17}.

In this work we make the following contributions to \ac{FDD} channel estimation via covariance identification in wideband massive \ac{MIMO} systems:
\begin{itemize}
	\item  We propose a wideband channel covariance identification method for \ac{FDD} massive \ac{MIMO} systems that can be utilized for both sub-$6$ GHz and \ac{mmWave} frequencies.
	\item With the aim of allowing for very short training sequences, Wichman sparse rulers \cite{Wic63} are used to design the channel estimation training sequences. 
	\item  The proposed method is applicable to a broad variety of scenarios because it only assumes 1) the channel is wide sense stationary over a period of time, and 2) the channel covariance matrix has a Toeplitz structure, a circumstance that occurs in usual antenna arrangements like \ac{ULA} and \ac{UPA}.
	\item We show that the proposed method supports the limitations imposed by the constraints of hybrid architectures. Moreover, it is robust against the beam squint effect.
	\item We propose a low complexity channel estimator which exploits the wideband structure of the channels to avoid the individual per-subcarrier estimation procedure. 
	
\end{itemize}

\section{System Model}

We consider a \ac{FDD} massive \ac{MIMO} system with $M$ transmit antennas communicating with single antenna users \cite{Fan17}. To overcome the frequency selective channel and avoid intercarrier interference, we assume an \ac{OFDM} modulation with a cyclic prefix of length $N$. We also assume that within each channel coherence block a portion of the block is dedicated to channel estimation while the remaining time is devoted to data transmission. More specifically, we assume that during the $k$-th channel coherence block the transmitter sends a frequency flat training sequence $\B{X}_k\in\mathbb{C}^{T_{\text{tr}}\times M}$ where $T_{\text{tr}}$ is the number of training channel uses.  In the frequency domain, the $k$-th received training block for the $\ell$-th subcarrier reads as
\begin{align}
\B{\varphi}_k[\ell] = \B{X}_k\B{h}[\ell]+ \B{v}[\ell],
\end{align}
where $\B{h}[\ell]\sim \mathcal{N}_\mathbb{C}(0, \B{C}_{\B{h}[\ell]})$ is the channel vector, and $\B{v}[\ell]\sim \mathcal{N}_\mathbb{C}(0, \sigma_{\B{v}[\ell]}^2\B{I}_{T_{\text{tr}}})$ is the noise, $k \in \{1, \ldots , K\}$ and $K$ being the total number of channel blocks.  Note that the number of training samples $T_{\text{tr}}$ is restricted due to the limited length of the channel coherence blocks, and this limits the number of possible orthogonal training sequences in $\B{X}_k$. However, the training symbols are vectors of size $M \gg T_{\text{tr}}$, due to the large antenna array deployed at the \ac{BS}. Hence, $\B{X}_k$  is a wide matrix, and the feasible training sequences are non orthogonal. In summary, the channel vector response of size $M$ at the $k$-th channel block and $\ell$-th subcarrier is estimated from the received vector $\B{\varphi}_k[\ell]$ of size $T_{\text{tr}}$. 

\begin{figure}[t]
	\includegraphics[width=.99\columnwidth]{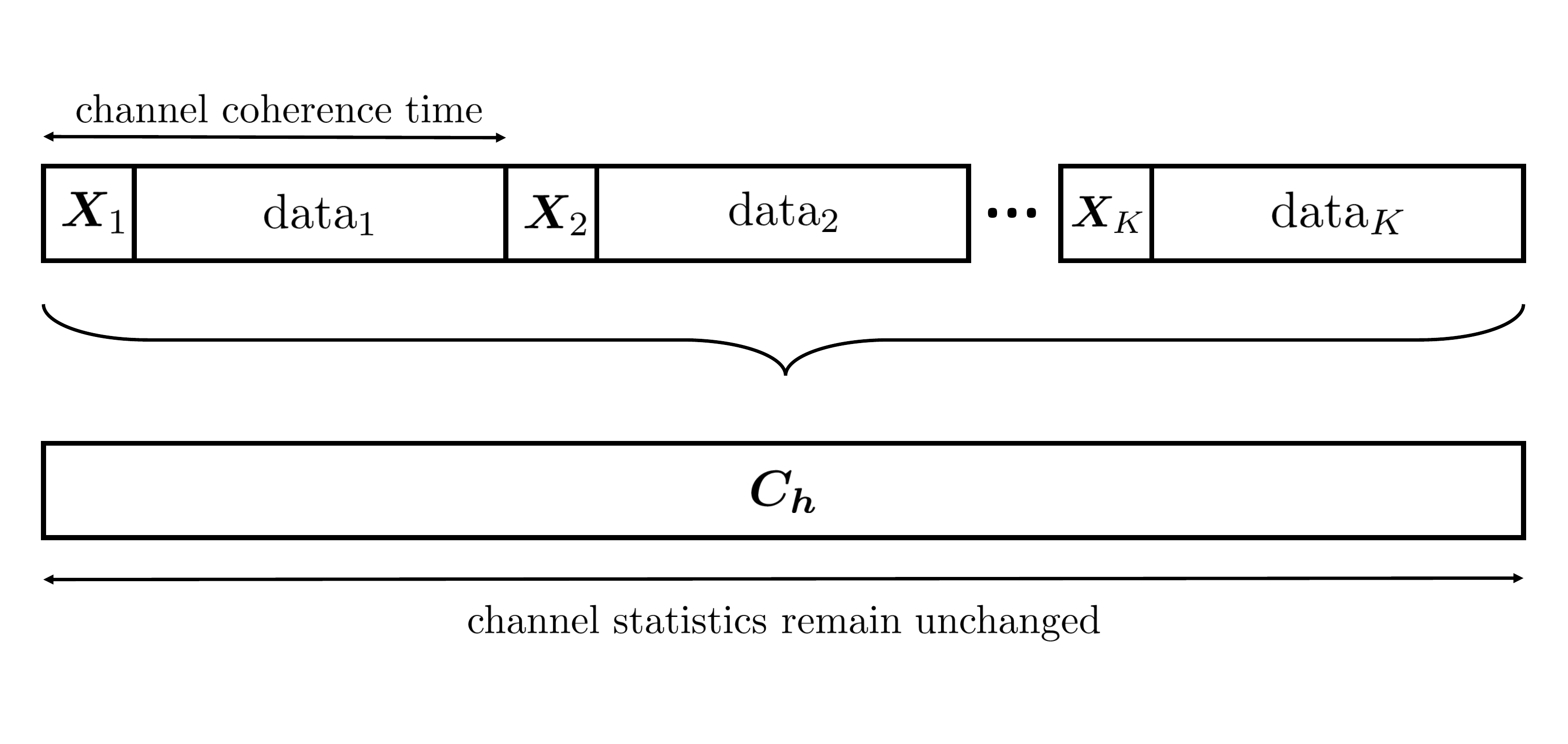}
	\caption{Channel and covariance coherence times.}
	\label{fig:times}
\end{figure}

Since we aim at determining the channel covariance, and assuming that $\B{h}[\ell]$ is stationary, we will consider several channel coherence blocks $k \in \{1, \ldots , K\}$. We now consider the specific realization of the received signal during the $k$-th training block and $\ell$-th subcarrier as
\begin{equation}
\B{\phi}_k[\ell] = \B{X}_k\B{h}_k[\ell]+ \B{v}_k[\ell].
\label{eq:receivSigRealiz}
\end{equation} 
As previously mentioned, the channel covariance slowly changes over time and its acquisition is thereby simpler than estimating the channel itself (see Fig. \ref{fig:times}). Since the channel and the noise vectors are statistically independent, the covariance of the observations is given by the expression 
\begin{align}
\B{C}_{\B{\varphi}_k[\ell]} = \B{X}_k\B{C}_{\B{h}[\ell]}\B{X}_k^H + \B{C}_{\B{v}[\ell]}.
\label{eq:observed_covariance}
\end{align}

Regarding the feedback, we assume that the users send the received signal back to the \ac{BS} to perform the channel covariance identification. This strategy was followed, e.g., in \cite{HuHuXuYazeroNorm}. Since we consider short training sequences, the amount of feedback data to be sent is usually smaller than that when the channel estimation is performed at the user end, and the users have to report the  positions of the non-zero elements and the corresponding gains \cite{FlRuTu18}. This scheme is especially useful when the number of channel paths $L$ is large.
\section{Channel Model}
We assume the channel response to an arbitrary user at the $n$-th delay tap reads as \cite{SchSa14,PaHe17,GoRoGoCaHe18} 
\begin{equation}
\B{h}_{n} = \sum_{l=1}^L  g_{l}p_{\text{rc}}(nT_s-\tau_l)\B{a}(\vartheta_{l}),
\label{eq:wideband_channel_model}
\end{equation}
where $g_{l}\sim\mathcal{N}_\mathbb{C}(0,\sigma_{l}^2)$ and $\tau_l$ are the complex channel gain and the relative delay corresponding to the $l$-th path, respectively, with $L$ the number of channel paths. We consider that the channel gains for different paths are statistically independent, i.e. $E[g_{l} g_{l'}^*] = \sigma_{l}^2\delta(l-l')$.  $p_{\text{rc}}(t)$ is the raised cosine pulse sampled with time interval $T_s$. Finally, we assume a \ac{ULA} at the transmitter so that the steering vectors $\B{a}(\vartheta_{l})$ have the form $\B{a}(\vartheta_{l})=[1, \eul^{-\imj \frac{2\pi}{\lambda} d \sin\vartheta_{l}},\ldots,\eul^{-\imj \frac{2\pi}{\lambda} d (M-1) \sin\vartheta_{l}} ]^T$, where $\vartheta_{l}$ is the \ac{AoD} for the $l$-th channel propagation path, $d$ is the distance between two consecutive array elements, and $\lambda$ represents carrier wavelength. Thus, the channel frequency response for the $\ell$-th subcarrier reads as
\begin{align}
\B{h}[\ell] &= \sum_{n=0}^{N-1} \B{h}_{n}\eul^{-\imj 2\pi \ell n/N_c}
=\sum_{l=1}^Lg_{l}\beta_l[\ell]\B{a}(\vartheta_{l})[\ell]\notag\\&=\B{A}[\ell]\B{\Theta}[\ell]\B{g},
\label{eq:wideChAngDomain}
\end{align}
where $N$ is the number of delay taps,  $$\beta_l[\ell]=\sum_{n=0}^{N-1}  p_{\text{rc}}(nT_s-\tau_l)\eul^{-\imj 2\pi \ell n/N_c },$$  $\B{\Theta}[\ell]=\operatorname{diag}(\beta_1[\ell],\ldots,\beta_L[\ell])$, and we remark the frequency dependency of the steering vectors with the subcarrier index $\ell$. That is, we consider the so-called beam squint effect that arises for large relative signal bandwidth $B/f_c$ \cite{Va02}. Thus, we define the relative position of the $N_c$ subcarriers by means of the vector $\B{\Delta}\in\mathbb{Z}^{N_c}$ as follows
\begin{align*}
\B{\Delta}^T=&\left[0,-1,\ldots,-\floor*{\frac{N_c}{2}},\floor*{\frac{N_c-1}{2}},\right.\\
&\left.\floor*{\frac{N_c-1}{2}}-1,\ldots,1\right]
\end{align*} 
with $\floor*{\cdot}$ the floor operator. Hence, for the central frequency $f_c$ the offset of the $\ell$-th subcarrier is $\Delta[\ell]\frac{B}{N_c}$. The matrix $\B{A}[\ell]$ is given by the Hadamard product $\B{A}\odot\B{\Upsilon}[\ell]$, with 
\begin{equation}
\B{\Upsilon}[\ell]=\begin{pmatrix}
1 & \ldots & 1\\
\eul^{\imj 2\pi d \sin\vartheta_{1}\Delta[\ell]} & \ldots & \eul^{\imj 2\pi d \sin\vartheta_{L}\Delta[\ell]}\\
\vdots &  & \vdots\\
\eul^{\imj 2\pi d \sin\vartheta_{1}\Delta[\ell](M-1)} & \ldots & \eul^{\imj 2\pi d \sin\vartheta_{L}\Delta[\ell](M-1)}
\end{pmatrix},
\label{eq:beamSquintMatrix}
\end{equation} 
and $\B{A}=[\B{a}(\vartheta_{1}), \ldots, \B{a}(\vartheta_{L})]\in\mathbb{C}^{M\times L}$. Moreover, the channel delay taps in \eqref{eq:wideband_channel_model} can be approximated by means of the steering vectors comprised in the dictionary $\B{\tilde{A}}=[\B{a}(\theta_{1}), \ldots, \B{a}(\theta_{G})]\in\mathbb{C}^{M\times G}$ of size $G$, that is, $\tilde{\B{h}}[\ell]\approx\B{h}[\ell]$ with
\begin{align}
\tilde{\B{h}}[\ell]=(\B{\tilde{A}}\odot \B{\tilde{\Upsilon}}[\ell])\B{\tilde{\Theta}}[\ell]\tilde{\B{g}},
\label{eq:vchannel_model}
\end{align}
where the random vector $\tilde{\B{g}}\in\mathbb{C}^G$ and matrix $\B{\tilde{\Theta}}[\ell]$ comprise the $L$ non-zero channel gains and shaping pulse samples, and $\B{\tilde{\Upsilon}}[\ell]$ extends $\B{\Upsilon}[\ell]$. Note that equality holds if the \ac{AoD} lie on the dictionary. Similarly $\tilde{\B{\varphi}}_k[\ell]\approx\B{\varphi}_k[\ell]$.

Given the channel model of \eqref{eq:wideChAngDomain}, the channel covariance matrix reads as
\begin{align}
\B{C}_{\B{h}[\ell]}  
&=(\B{A}\odot\B{\Upsilon}[\ell])\B{\Theta}[\ell]\B{D}\B{\Theta}^H[\ell](\B{A}\odot\B{\Upsilon}[\ell])^H,
\label{eq:channel_covariance_wide}
\end{align}
where the diagonal matrix $\B{D}=\operatorname{diag}(\sigma_{1}^2,\ldots,\sigma_{L}^2)$ contains the channel gain variances. Furthermore, using \eqref{eq:vchannel_model}, the channel covariance matrix is approximated by 
\begin{equation}
\label{eq:vchannel_cov}
\B{C}_{\tilde{\B{h}}[\ell]}= (\B{\tilde{A}}\odot\B{\tilde{\Upsilon}}[\ell])\B{\tilde{\Theta}}[\ell]\B{\tilde{D}}\B{\tilde{\Theta}}^H[\ell](\B{\tilde{A}}\odot\B{\tilde{\Upsilon}}[\ell])^H,
\end{equation}
where $\B{\tilde{D}}=\operatorname{diag}(\tilde{\sigma}_1^2,\ldots,\tilde{\sigma}_G^2)$. Based on the latter approximation, we introduce  $\B{C}_{\tilde{\B{\varphi}}_k[\ell]} \approx\B{C}_{\B{\varphi}_k[\ell]}$ as
\begin{align}
\B{C}_{\tilde{\B{\varphi}}_k[\ell]} 
&=\B{\Psi}_k[\ell]\B{\tilde{D}}[\ell]\B{\Psi}_k^H[\ell]+ \B{C}_{\B{v}[\ell]},
\label{eq:vobserved_covariance}
\end{align}
with measurement matrices $\B{\Psi}_k[\ell]=[\B{\psi}_{1,k}[\ell],\ldots,\B{\psi}_{G,k}[\ell]]=\B{X}_k[\B{a}(\theta_{1})[\ell],\ldots,\B{a}(\theta_{G})[\ell]]$. According to this definition and assuming that both $\B{\Psi}_k[\ell]$ and $\B{C}_{\B{v}[\ell]}$ are known, identifying $\B{C}_{\tilde{\B{\varphi}}_k[\ell]}$ is equivalent to estimating $\B{\tilde{D}}[\ell]=\B{\tilde{\Theta}}[\ell]\B{\tilde{D}}\B{\tilde{\Theta}}^H[\ell]$.

\section{Estimation Scheme and Training Design}
We start summarizing the proposed scheme
\begin{itemize}
	\item At each time slot, the \ac{BS} transmits a predefined training sequence to the users, and the users feed the observations back to the \ac{BS} using, e.g., scalar quantization \cite{GaDaWaCh15,HuHuXuYa17}.
	\item In the preamble, the channel covariance matrices are estimated for all the users and frequency bands. 
	\item Due to the slow variation of the channel statistics, the \ac{BS} keeps accurately tracking the channel covariance matrices using subsequent observations.
	\item Since the training sequence is the same for covariance identification and channel estimation, the observations are also employed to estimate the channel responses within the coherence time taking into account the estimated channel covariance.
\end{itemize}

Covariance matrices of size $M\times M$ have $(M^2 + M)/2$ different coefficients. A training sequence of length $T_{\text{Tr}}$ allows to define $(T_{\text{Tr}}^2 + T_{\text{Tr}})/2$ equations. If a channel covariance matrix is Toeplitz, the number of matrix coefficients to estimate reduces to $M$, and this establishes the following trivial lower bound on the training length 
\begin{align}
T_{\text{Tr}} \ge \sqrt{2M+1/4}-1/2.
\end{align}
A necessary condition to achieve this bound is that the training sequence should allow us to measure the correlation between pairs of antennas for any separation. This has a strong connection with sparse complete rulers, defined mathematically as an ordered sequence of integers $\mathcal{R}= (0,r_1,\ldots, r_T)$ such that any $z\le r_T$ can be written as $z = r_i - r_j$ for some $r_i$ and $r_j$ in $\mathcal{R}$. In this case, we speak of a complete sparse ruler of length $r_T$, and a perfect ruler is a complete ruler that contains the minimum possible number of elements for a given $r_T$.

It has been shown that a training sequence built from a perfect sparse ruler like $\B{X} = [\B{e}_{r_1},\ldots,\B{e}_{r_T}]^T$, with $\B{e}_i$ the indicator vector of zeros and a one in the $i$-th element, is sufficient for determining the covariance matrix \cite{RomLeus13}, and they were applied to the covariance identification problem \cite{ShArLe12}. In the following, we summarize some results found in the literature for sparse rulers. First, there exist bounds on the size of perfect rulers for a given length \cite{Lee56}.

\begin{theorem}
(see \cite{Lee56}) Denoting as $l(r)$ the minimum number of marks of a perfect sparse ruler of length $r$, then the $\lim\limits_{r \rightarrow \infty} \frac{l^2(r)}{r}$ exists and
\begin{align}
2.434 \le \lim\limits_{r \rightarrow \infty} \frac{l^2(r)}{r} \le \frac{375}{112}\approx3.348.\label{eq:bound_complete_ruler}
\end{align}
\end{theorem}
Hence, we can establish that it is possible to identify a Toeplitz covariance matrix, using sparse rulers, with a training length that satisfies $\sqrt{2.434M}\le T_{\text{tr}}\le\sqrt{3.348M}$ when $M$ approaches infinity. 

If we drop the constraint that the ruler must be perfect, i.e. that the largest element of the ruler must be equal to the length of the ruler $r_T=r$, tighter upper bounds can be established. This result is interesting for cases when it is possible to assume that the covariance matrix is banded.
\begin{theorem}
	(see \cite{Lee56}) Denoting as $k(r)$ the minimum number of marks of a sparse ruler of length $r$, then $\lim\limits_{r \rightarrow \infty} \frac{k^2(r)}{r}$ exists and
	\begin{align}
	2.434 \le \lim\limits_{r \rightarrow \infty} \frac{k^2(r)}{r} \le \frac{8}{3}\approx 2.6571.\label{eq:bound_incomplete_ruler}
	\end{align}
\end{theorem}
This implies that shorter training lengths can be achieved for the same number of coefficients if the covariance matrix is banded.


A trivial example of sparse ruler can be found for any given number $m$, since it is always possible to build a perfect sparse ruler of length $m^2$ with elements $(0, 1,2,\ldots,m,2m,3m,\ldots,m^2)$. This provides a length of $m^2$ different differences with $2m$ elements, thus achieving a value of $\lim\limits_{r \rightarrow \infty}\frac{l^2(r)}{r}=4$, larger than the upper bound established above. Some works on \ac{AoD} estimation also use coprime arrays \cite{PaVa11,HaCa18}. Moreover, this method is based on two co-prime numbers $p$, $q$, i.e. $\operatorname{mcd}(p,q)=1$, and on generating the ruler as the set $\{mp:0\le m < q \} \cup \{ nq:0\le n < p  \}$. This method achieves a training length of $T_{\text{tr}} = p+q$ and allows to identify $pq$ different lags. Considering the limiting function $\frac{l^2(r)}{r}$ we obtain $\frac{(p+q)^2}{pq}$, and approximating $p\approx q \approx \sqrt{M}$ it causes the limit in \eqref{eq:bound_complete_ruler} to be $4$ when the number of antennas grows to infinity. Hence, this strategy cannot improve the asymptotic efficiency of the previous trivial ruler.

In the literature we can find some techniques to build sparse rulers for arbitrary lengths satisfying the bound in \eqref{eq:bound_incomplete_ruler}, as summarized in Sec. \ref{sec:diffSetRulers} and \ref{sec:Wichmann}.

\subsection{Difference Set-based Rulers}
\label{sec:diffSetRulers}

A perfect difference set $\mathcal{S} = (0,s_1,\ldots,s_q)$ respect to $p$ is a set of integers that contains all differences modulo $p$ exactly once, i.e. $a,b,a',b'\in \mathcal{S}$ and $a-b=a'-b'=v$, then $a=a'$ and $b=b'$. The first example of such sets and a systematic procedure to find them was discovered by Singer. In particular, he proved that for any prime number $q$ there exists a perfect difference set respect to $q^2+q+1$ with $q+1$ integers \cite{Sin38}.

Any perfect difference set in combination with other sparse ruler can be employed to build sparse rulers of arbitrarily large lengths. In particular, if $\mathcal{R} = (0,r_1, \ldots, r_T)$ is a sparse ruler of length $r_T$ with $T+1$ elements, a new sparse ruler can be defined with the help of a perfect difference set $\mathcal{S}$ respect to $m=q^2+q+1$ as $\mathcal{T}=(0,t_1, \ldots, t_{(T+1)(q+1)})$ with elements of the form
\begin{align}
t_k=r_im+s_j, r_i\in\mathcal{R}, s_j\in\mathcal{S}.
\end{align}
In this case, the ruler $\mathcal{T}$ contains all differences between $1$ and $R=r_Tm+m-s_q-1$ \cite{Lee56}. However, from $R+1$ to the largest difference in $\mathcal{T}$, $r_Tm+s_q$, there are some missing differences. Leech proved that this set can be completed to contain all differences for any $R\ge r_Tm+s_q$ by adding at most $2 q$ elements, with $q$ the prime number that generated the difference perfect set $\mathcal{S}$ \cite{Lee56}. This relies on the following observation: it is always possible to build a difference set that contains all differences in the interval $[r,s]$ with a set of the form $(0,1,2,\ldots,t,r+t,r+2t+1,\ldots,s)$, with $t=\lfloor\sqrt{s-r}\rfloor$. The cardinality of this set is at most $2t+2$. Following this procedure, it is possible to complete the ruler from length $r_Tm+m-s_q-1$ to $r_Tm+s_q$, building a new ruler from the union of both sets.

\subsection{Wichmann Rulers}
\label{sec:Wichmann}
Wichmann proposed a direct method to obtain sparse rulers of any size that, denoted as the differences between consecutive terms, have the form
\begin{align}
\underbrace{1,\ldots,1}_r,r+1,\underbrace{2r+1,\ldots,2r+1}_{r},\underbrace{4r+3,\ldots,4r+3}_{s},\notag\\
\underbrace{2r+2,\ldots,2r+2}_{r+1},\underbrace{1,\ldots,1}_{r}
\end{align}
for some parameters $r$ and $s$ such that $2r-2\le s\le 2r+4$ \cite{Wic63}. This allows to find the values for $r$ and $s$ that generate a sparse ruler with a length as close as possible to the desired length. Finally, the ruler can be completed up to the desired length following the same approach as in the previous section.

\subsection{Hybrid Architectures}
Hybrid architectures improve the hardware efficiency and power consumption costs by reducing the number of \ac{RF} chains $N_\text{RF}$. The counterparts are: 1) the lack of flexibility of the analog processing network, typically complex weights of fixed magnitude that are flat in the frequency domain; and 2) the dimensionality reduction because of the relationship $N_\text{RF}\ll M$. Recall that we consider the scenario with single antenna users. Thus, when the training is performed in the uplink, the received signal is affected by the analog receive filter. This filter is frequency flat and has to be designed without channel knowledge. These two facts lead to longer training overheads. On the contrary, with the proposed downlink training strategy,  only two \ac{RF} chains are enough to achieve the performance of fully digital architectures. This is because of the utilization of frequency flat training vectors based on sparse rulers. Let us denote the analog and digital precoding matrices as $\B{P}_{\text{A},k}(t)\in\mathbb{C}^{M\times N_\text{RF}}$ and $\B{p}_\text{D}(t)\in\mathbb{C}^{N_\text{RF}}$, such that $\B{w}_k(t)=\B{P}_{\text{A},k}(t)\B{p}_\text{D}(t)$ is the training sequence using a hybrid architecture. Consider the training sample $t$ at the $k$-th training block $\B{x}_{k}(t)$. For $N_\text{RF}=2$ we have $\B{w}_k(t)=\B{x}_{k}(t)$ by setting
\begin{equation}
[\B{P}_{\text{A},k}(t)]_{i,:}=\begin{cases}
\begin{bmatrix}
\eul^{\imj\phi} & \eul^{\imj(\phi+\pi)}
\end{bmatrix},& \text{if } [\B{x}_{k}(t)]_i = 1\vspace*{0.1cm}\\
\begin{bmatrix}
\eul^{\imj\phi} & \eul^{\imj\phi}
\end{bmatrix},              & \text{otherwise,}
\end{cases}
\end{equation}
with $\B{p}_\text{D}=\frac{1}{2}\B{1}$, and the all ones vector $\B{1}$. Note that the phase can be easily compensated, e.g., with the training sequence. 

Recently, the advantages of dynamic time-variant training sequences for hybrid architectures have been analyzed\cite{PaHe17}. The proposed channel estimation method is valid for this setup by using different sparse rulers of a given length.

\section{Covariance Matrix Identification}
In this section we present and analyze different algorithms to perform covariance identification in wideband massive \ac{MIMO} systems. 

\subsection{ML Estimator}
\ac{ML} is the estimator maximizing the \ac{LLF}. Typical covariance identification strategies employ all the observations to estimate the gain variances $\B{\tilde{D}}[\ell]$. In our scenario, however, these variances are not common for the different subcarriers disabling a trivial extension that fully exploits the common information among all frequencies. Thus, in this section we drop the subcarrier index $\ell$ and define the \ac{LLF} for the observation $\B{\varphi}_k$ as
\begin{align}
L \left(\B{\varphi}_k; \B{\tilde{D}} \right) 
 =&\log\det \left(\B{C}_{\B{\varphi}_k} \right) 
  + \operatorname{tr}\left(\B{C}_{\B{\varphi}_k}^{-1} \B{\varphi}_k\B{\varphi}^H_k\right). \label{eq:log_likelihood2}
\end{align}
Assuming now independent $\B{\varphi}_k$, the joint \ac{LLF} is $L(\B{\varphi}_1,\ldots,\B{\varphi}_K; \B{\tilde{D}}) = \sum_{k=1}^{K} L(\B{\varphi}_k; \B{\tilde{D}})$.

The ML estimate is obtained by solving
\begin{align}
\label{eq:ml_problem}
\underset{\B{\tilde{D}}}{\min}&~~L(\B{\phi}_1,\ldots,\B{\phi}_K; \B{\tilde{D}}) ~~s.t.~~\tilde{\sigma}_g^2\ge 0, \forall g\in [1,G]. 
\end{align}
When the training blocks are the same for all periods,  $\B{C}_{\B{\varphi}_k} = \B{C}_{\B{\varphi}}, \forall k$, and hence the log-likelihood function \eqref{eq:log_likelihood2} for a block of $K$ snapshots simplifies to
\begin{align*}
\frac{1}{K}L(\B{\varphi}_1,\ldots,\B{\varphi}_K; \B{\tilde{D}}) = \log\det \left( \B{C}_{\B{\varphi}} \right) 
+ \operatorname{tr}\left(\B{C}_{\B{\varphi}}^{-1}\hat{\B{C}}_{\B{\varphi}} \right),
\end{align*}
where $\hat{\B{C}}_{\B{\varphi}}= \frac{1}{K}\sum_{k=1}^{K}\B{\varphi}_k\B{\varphi}_k^H$ is the sample estimate covariance matrix  \cite{Rom13}. 
Even for the latter \ac{LLF}, the problem of \eqref{eq:ml_problem} is tough since it involves the sum of a concave and a convex function. Multiple algorithms have been proposed to tackle \eqref{eq:ml_problem} \cite{Rom13}, e.g. LIKES \cite{Bab12}. This approach is based on the majorization-minimization principle and offers a robust solution at the expense of a large computational burden.

\subsection{Considerations for wideband approaches}
It is apparent that per-subcarrier (or per-sub-band) covariance identification is not a practical approach. First, it does not take advantage of the frequency flat nature of $\B{D}$ in \eqref{eq:channel_covariance_wide}, and it does not leverage the symmetry with respect to the central frequency. Second, system scalability of this strategy may be compromised since wideband dictionaries have to be computed for each iteration or stored. Finally, the larger number of subcarriers, the more covariance matrices to identify.
Hence, we propose a joint recovery of the support for all the subcarriers, and reducing almost by a half the number of gain variances to be estimated. 

To that end, let us start by highlighting some observations that are key for the proposed algorithms. 

1) We write the wideband steering vectors as
\begin{equation}
\B{a}(\vartheta_{l})[\ell]=\B{\Gamma}(\vartheta_{l})[\ell]\B{a}(\vartheta_{l}),	
\label{eq:steering_BS}
\end{equation} 
where the beam-squint effect can be modeled with the matrix
\begin{equation}
\B{\Gamma}(\vartheta_{l})[\ell]=\operatorname{diag}(1,\eul^{\imj 2\pi d \sin\vartheta_{l}\Delta[\ell]},\ldots,\eul^{\imj 2\pi d \sin\vartheta_{l}\Delta[\ell](M-1)}).
\label{eq:steeringWideband}
\end{equation}

2) Recall that the path delays $\tau_l$ of \eqref{eq:wideband_channel_model} are unknown, as well as the matrices $\B{\tilde{\Theta}}[\ell]$ in \eqref{eq:vchannel_cov}. Therefore, we aim at estimating the per-carrier gain variances $\B{\tilde{D}}[\ell]$. 

3) Regarding the channel gains, for subcarriers $\ell$ and $\ell^\prime$ sharing the same frequency offset, i.e. $|\Delta[\ell]|=|\Delta[\ell^\prime]|$, $\B{D}[\ell]=\B{D}[\ell^\prime]$ holds. Recall that each diagonal element of $\B{D}_{l,l}[\ell]$ can be written as $\beta_l[\ell]\B{D}_{l,l}\beta_l^*[\ell]$. Then, thanks to the symmetry of the Fourier transform for real signals we have that $\beta_l[\ell]=\beta_l^*[\ell']$. In particular,
\begin{align}
\label{eq:symmFour}
\beta_l[\ell]&=\sum_{n=0}^{N-1}  p_{\text{rc}}(nT_s-\tau_l)\left(\eul^{-\imj 2\pi(-\ell) n/N_c }\right)^*,
\end{align}
or  $\beta_l[\ell]=\beta_l^*[-\ell\operatorname{mod}N_c]$, thus leading to $\ell'=-\ell\operatorname{mod}N_c$.

In the following we develop some methods that incorporate the full channel structure into the identification process.

%

\subsection{Wideband OMP-based estimator}
Assuming that $\B{\tilde{D}}[\ell]\in\mathcal{H}$, with $\mathcal{H}$ the set of $L$-sparse hermitian matrices, the \ac{OMP} algorithm can be adapted to the quadratic case to find an approximate solution to the non-convex problem
\begin{align}
\hat{\B{D}}[\ell]=\underset{\B{\tilde{D}}[\ell]\in\mathcal{H}}{\min}~~ & \left\Vert \hat{\B{C}}_{\B{\varphi}_k[\ell]} - \B{\Psi}_k[\ell]\B{\tilde{D}}[\ell]\B{\Psi}_k^H[\ell] \right\Vert^2_F,
\end{align}
by employing \ac{COMP} and \ac{DCOMP} algorithms \cite{PaHe17}. These algorithms greedily determine the covariance matrix support  $\mathcal{S}$ that contains the integer indices $\mathcal{S}=\{s_1,\ldots,s_L\}\subseteq \{1,\ldots,G\}$. Nevertheless, the extension of \ac{DCOMP} to the wideband scenario in \cite{PaHe17} neglects the beam-squint effect. 

In order to consider a more general scenario, we propose to use the following criterion for the $n$-th iteration 
\begin{equation}
s_n=\underset{j\in\{1,\ldots,G\}}{\max} \sum_{\ell=1}^{N_c}\sum_{k=1}^K  \B{\psi}_{j,k}[\ell]\B{R}_{n-1,k}[\ell]\B{\psi}_{j,k}^H[\ell],
\label{eq:WCOMPcriterion}
\end{equation}
where
\begin{equation}
\B{\psi}_{j,k}[\ell]=\B{X}_k\B{\Gamma}(\theta_{j})[\ell]\B{a}(\theta_{j})
\label{eq:phiWide},
\end{equation}
$\mathcal{S}$ is updated as $\mathcal{S}=\mathcal{S}\cup s_n$, and the wideband residual is given by $\B{R}_{n-1,k}[\ell]=\hat{\B{C}}_{\B{\varphi}_k[\ell]}-\B{\Psi}_{\mathcal{S},k}[\ell]\B{\Psi}_{\mathcal{S},k}^{\dag}[\ell]\hat{\B{C}}_{\B{\varphi}_k[\ell]} (\B{\Psi}_{\mathcal{S},k}[\ell]\B{\Psi}_{\mathcal{S},k}^{\dag}[\ell])^H$, with  $\B{\Psi}_{\mathcal{S},k}[\ell]=[\B{\psi}_{s_1,k}[\ell],\ldots,\B{\psi}_{s_n,k}[\ell]]$. Note that $\B{\psi}_{j,k}[\ell]$ 
is defined taking into account the offset with respect to the central frequency $f_c$ by means of the multiplication times $\B{\Gamma}(\theta_{j})[\ell]$ in \eqref{eq:steeringWideband}. 
This way we find the support jointly for all the  subcarriers, $\B{\Psi}_{\mathcal{S},k}[\ell],\,\forall \ell$. After $L$ steps, the channel gain variances are estimated for each subcarrier by computing
\begin{align}
\hat{\B{D}}_{\mathcal{S}}[\ell] =&\frac{1}{K}\sum_{k=1}^K\B{\Psi}_{\mathcal{S},k}^{\dag}[\ell]\hat{\B{C}}_{\B{\varphi}_k[\ell]} (\B{\Psi}_{\mathcal{S},k}^H[\ell])^{\dag},
\label{eq:WCOMPestim}
\end{align}
where $\hat{\B{D}}_{\mathcal{S}}$ contains the gains associated to the vectors in $\B{\Psi}_{\mathcal{S},k}$.  Finally, $\hat{\B{D}}[\ell]$ is computed as a matrix of zeros except for elements at the rows and columns selected by $\mathcal{S}$.

Notice that this algorithm allows to change the training sequence for each period, and $\B{C}_{\B{\varphi}_k[\ell]}$ is then a one sample estimate of the sample
covariance matrix. This method improves the estimation quality when hybrid architectures are considered for uplink training \cite{PaHe17}. The use of the one sample estimates tough, disallows to exploit the fact that $\B{\tilde{D}}[\ell]$ is diagonal and increases the computational complexity of channel estimation, see Sec. \ref{sec:complexity}.

\subsection{MUSIC}
\ac{MUSIC} is a well-known estimation algorithm which is based on the orthogonality of signal and noise subspaces. In our particular scenario, the \ac{MUSIC} algorithm can be employed to identify the \ac{AoD} corresponding to the $L$ paths if the training sequences are constant, i.e. $\B{X}_k = \B{X}, \forall k$. 

The first step of \ac{MUSIC} consists in computing the eigendecomposition of the sample covariance matrix $\hat{\B{C}}_{\B{\varphi}[\ell]}=\B{U}[\ell]\B{\Lambda}[\ell]\B{U}^H[\ell]$, assuming that the covariance converges to the second order moment of the channel for a large number of samples, that is, $\hat{\B{C}}_{\B{\varphi}[\ell]} \rightarrow \B{C}_{\B{\varphi}[\ell]}$ when $K\rightarrow\infty$. Next, the eigenvectors corresponding to the $L$ largest eigenvalues are discarded to obtain the basis $\bar{\B{U}}[\ell]$, that spans the noise subspace. Due to the beam-squint effect, the relationship among dictionary matrices for different frequencies is non-linear. Therefore, it is not feasible to perform this step jointly for all the subcarriers unless the relative signal bandwidth $B/f_c$ is very small. A remarkable feature of this subspace discrimination is its lack of dependency with the SNR, as long as the noise is spatially white. Under such assumption, for the approximation in \eqref{eq:vobserved_covariance} we get
\begin{align}
\label{eq:MUSICeigen}
	\B{C}_{\tilde{\B{\varphi}}[\ell]}&= \B{\Psi}[\ell]\B{\tilde{D}}[\ell]\B{\Psi}^H[\ell] + \sigma_{\B{v}[\ell]}^2\mathbf{I}_{T_\text{TR}}\notag\\
	&=\B{U}[\ell](\B{\Lambda}[\ell]+\sigma_{\B{v}[\ell]}^2\mathbf{I}_{T_\text{TR}})\B{U}^H[\ell].
\end{align}
Therefore, it is crucial to force this condition by applying a pre-whitening filter if necessary. According to this observation, the search of the support is performed in the effective subspace, contrarily to other methods like \ac{COMP}. 

The second step of \ac{MUSIC} performs the identification of the support $\mathcal{S}$.  Using the steering vectors for different frequencies of \eqref{eq:steering_BS}, we build the following joint estimator function for all the subcarriers, 
\begin{equation}
\mathcal{J}_i=\sum_{\ell=1}^{N_c}\frac{1}{\|\B{\psi}_{i}^H[\ell]\bar{\B{U}}[\ell]\|_2^2},
\label{eq:musicWide}
\end{equation} 
with the definition of $\B{\psi}_{i,k}[\ell]$ in \eqref{eq:phiWide} for $\B{X}_k=\B{X},\,\forall k$, to increase the robustness of the decision. Thus, the angles are identified as the  $L$ larger values of $\mathcal{J}_i$.

Once the angles are determined, we construct the matrices $\B{\Psi}_{\mathcal{S}}[\ell] =\B{X} [\B{\Gamma}(\theta_{s_1})[\ell]\B{a}(\theta_{s_1}),\ldots,\B{\Gamma}(\theta_{s_L})[\ell]\B{a}(\theta_{s_L})]$ with $\mathcal{S}$ the set of $L$ indices corresponding to the largest values of $ \mathcal{J}_i$. To recover the associated gain variances $\B{\hat{D}}_{\mathcal{S}}[\ell]$, it is possible to use the estimates in \cite{Sc86}, i.e.,
\begin{equation}
	\B{\hat{D}}_{\mathcal{S}}[\ell]=\B{\Psi}_{\mathcal{S}}^\dagger[\ell](\hat{\B{C}}_{\B{\varphi}[\ell]}-\sigma_{\B{v}[\ell]}^2\mathbf{I})(\B{\Psi}_{\mathcal{S}}^H[\ell])^\dagger.
	\label{eq:LS_fullrank}
\end{equation}
This direct solution clearly depends on the rank of $\B{\Psi}_{\mathcal{S}}[\ell]$, and cannot be applied in the following situations
\begin{enumerate}
	\item $L> T_\text{tr}$. This scenario arises for very short training sequences, or scenarios with a large number of propagation paths. Under this assumption the pseudo-inverse $\B{\Psi}_{\mathcal{S}}^\dagger[\ell]$ does not exist and \eqref{eq:LS_fullrank} cannot be computed.
	\item $L \leq T_\text{tr}$. In this case the rank of $\B{\Psi}_{\mathcal{S}}[\ell]$ depends on the choice of the dictionary matrix $\B{\tilde{A}}$, for proper training matrix $\B{X}$. Recall that, the dictionary contains non-orthogonal vectors when $M$ is finite. When the distance between two consecutive angles in the dictionary is small, a situation that arises when $G$ is large, this may lead to dictionary matrices such that $\operatorname{rank}(\B{\tilde{A}})\leq\min\{M,G\}$, thus making $\B{\Psi}_\mathcal{S}^H[\ell]\B{\Psi}_\mathcal{S}[\ell]$ ill-conditioned.
\end{enumerate}

To circumvent these situations, one possibility is to reduce the dictionary size, see Sec. \ref{sec:DicSizeLEstim}. Another approach consists on employing \ac{OSMP} to select the steering vectors, hence avoiding linear dependencies \cite{Lee12,TsLiWu18}. Alternatively, we propose to exploit the diagonal structure of $\B{\tilde{D}}[\ell]$, and perform the estimation by means of
\begin{equation}
	\operatorname{vec}(\B{\hat{D}}_{\mathcal{S}}[\ell])=(\B{\Psi}_{\mathcal{S}}^*[\ell]\circ \B{\Psi}_{\mathcal{S}}[\ell])^\dag\operatorname{vec}(\hat{\B{C}}_{\B{\varphi}[\ell]}-\sigma_{\B{v}[\ell]}^2\mathbf{I}_{T_\text{TR}}),
	\label{eq:gainEstimates}
\end{equation}
with $\circ$ the Khatri–Rao product (defined as the column-wise Kronecker product). The Khatri–Rao product produces matrices of rank $L$ when the columns of  $\B{\Psi}_{\mathcal{S}}[\ell]$ are not co-linear. To prove the last statement we consider two cases:

1) If the columns of $\B{\Psi}_{\mathcal{S}}[\ell]$ are linearly independent, the same property holds for the columns of $\B{\Psi}_{\mathcal{S}}^*[\ell]\circ \B{\Psi}_{\mathcal{S}}[\ell]$. 

2)  Consider that the $i$-th column of $\B{\Psi}_{\mathcal{S}}[\ell]$, $\B{\psi}_{\mathcal{S}_i}$ for notation simplicity, is a linear combination of the columns whose indices are in the set $\mathcal{T}\subseteq\mathcal{S}$, i.e.,
$\B{\psi}_{\mathcal{S}_i}=\sum_{j=1}^{|\mathcal{T}|}c_j\B{\psi}_{\mathcal{T}_j}$ with $\B{\psi}_{\mathcal{T}_j}\neq \mathbf{0},\forall j\in\mathcal{T}$. Then,  for the $i$-th column of $(\B{\Psi}_{\mathcal{S}}^*[\ell]\circ \B{\Psi}_{\mathcal{S}}[\ell])$ we obtain
\begin{equation}
\sum_{j=1}^{|\mathcal{T}|}c_j^*\B{\psi}_{\mathcal{T}_j}^*\otimes	\sum_{n=1}^{|\mathcal{T}|}c_n\B{\psi}_{\mathcal{T}_n}=\sum_{j=1}^{|\mathcal{T}|}\sum_{n=1}^{|\mathcal{T}|}c_j^*\B{\psi}_{\mathcal{T}_j}^*\otimes c_n\B{\psi}_{\mathcal{T}_n}.
\end{equation}
Notice that it is not possible to write this column as a linear combination of the form $\sum_{j=1}^{|\mathcal{T}|}b_j\B{\psi}_{\mathcal{T}_j}^*\otimes \B{\psi}_{\mathcal{T}_j}$ unless $\B{\psi}_{\mathcal{T}_j}=\sum_{n=1}^{|\mathcal{T}|}c_n\B{\psi}_{\mathcal{T}_n},\,\forall j$. This condition holds for $\B{\psi}_{\mathcal{S}_i}=c_j\B{\psi}_{\mathcal{T}_j}$.   

Then, resorting to the symmetry of \eqref{eq:symmFour} for $0<\ell<N_c/2$, we increase the robustness of the estimated gain variances with
\begin{align}
\label{eq:averagingGains}
\operatorname{vec}(\B{\hat{D}}_{\mathcal{S}}[\ell])&=\frac{1}{2}\big((\B{\Psi}_{\mathcal{S}}^*[\ell]\circ \B{\Psi}_{\mathcal{S}}[\ell])^\dag\operatorname{vec}(\hat{\B{C}}_{\B{\varphi}[\ell]}-\sigma_{\B{v}[\ell]}^2\mathbf{I})\\
&+(\B{\Psi}_{\mathcal{S}}^*[\ell']\circ \B{\Psi}_{\mathcal{S}}[\ell'])^\dag\operatorname{vec}(\hat{\B{C}}_{\B{\varphi}[\ell']}-\sigma_{\B{v}[\ell']}^2\mathbf{I})\big),\notag
\end{align}
where $\ell'=-\ell\operatorname{mod}N_c$. Observe that not all the gains can be calculated by means of the latter expression. In the case of odd  $N_c$, $\ell=0$ has to be determined using \eqref{eq:gainEstimates}, whereas $\ell=0$ and $\ell=N_c/2$ have to be computed via \eqref{eq:gainEstimates} for $N_c$ even.

Note that \ac{MUSIC} algorithm only works if the sample covariance has a rank of at least the number of elements to estimate. That is, the number of paths (or angles) estimated is bounded by the number of snapshots and the number of observations, i.e., $L\leq \min\{T_\text{tr},K\}$. Improvements to \ac{MUSIC} were proposed to overcome the rank deficient scenarios in  \cite{Lee12,Kim12}. In particular, these solutions apply when the number of snapshots $K$ is small or when they are linearly dependent, that is, when $\operatorname{rank}(\hat{\B{C}}_{\B{\varphi}}[\ell])<L$. Nevertheless, in the proposed scenario, it is desirable to reduce the training sequence length $T_\text{tr}$ as much as possible. Unfortunately, we obtain a more involved setup for which $$\operatorname{rank}(\hat{\B{C}}_{\B{\varphi}})=T_\text{tr},$$ with $T_\text{tr}<L$. This scenario can be addressed by using spatial smoothing, as described in the ensuing section.

\subsection{Spatial Smoothing}
\label{subsec:smooth}
\ac{SS} is an array processing technique based on nested arrays with different antenna separation \cite{PaVa10}. An effect similar to that of the nested arrays is obtained with the use of sparse rulers \cite{ArLe13}, which is equivalent to deploying non-uniform distance antenna elements. Applied to our scenario, the training sequence is built based on a perfect sparse ruler as $\B{X} = [\B{e}_{r_1},\ldots,\B{e}_{r_T}]^T$.
Thus, with the vectorization of the sample covariance matrix, $\B{y}[\ell]=\operatorname{vec}(\hat{\B{C}}_{\B{\varphi}[\ell]})$, we get 
\begin{align}
\B{y}[\ell] \approx (\B{\Psi}^*[\ell] \circ \B{\Psi}[\ell])\operatorname{diag}(\B{\tilde{D}}[\ell]) + \sigma_{\B{v}[\ell]}^2 \B{e}^\prime,\label{eq:vectorized_cov}
\end{align}
where the approximation comes from  \eqref{eq:vobserved_covariance}, and $\B{e}^\prime=[\B{e}_1^T,\ldots,\B{e}_{T_\text{tr}}^T]^T$. Observe that $(\B{\Psi}^*[\ell] \circ \B{\Psi}[\ell])$  contains the products  corresponding to the spatial differences $z\in\{-M+1,M-1\}$. \ac{SS} discards the repeated distances and sorts the remaining ones in a matrix $\B{B}[\ell]$ containing $2M-1$ rows from $(\B{\Psi}^*[\ell] \circ \B{\Psi}[\ell])$. The reduced vector is then
\begin{equation}
\check{\B{y}}[\ell]\approx\B{B}[\ell]\operatorname{diag}(\B{\tilde{D}}[\ell])+\sigma_{\B{v}[\ell]}^2\B{e}_M.
\label{eq:smoothing}
\end{equation}
The $2M-1$ differences in the reduced vector $\check{\B{y}}[\ell]\in\mathbb{C}^{2M-1}$ are next seen as a phase shift of the $M$ differences to be estimated. Considering $M$ overlapping subarrays, such that the $m$-th subarray  comprises the differences $\{-M+m,m-1\}$ in   $\check{\B{y}}_{m}[\ell]\in\mathbb{C}^{M}$, the spatial smoothed matrix is obtained averaging over the $M$ subarrays as follows
\begin{equation}
	\check{\B{Y}}[\ell]=\frac{1}{M}\sum_{m=1}^M\check{\B{y}}_{m}[\ell]\check{\B{y}}_{m}^H[\ell].
\end{equation}
As shown in \cite{PaVa10}, $\check{\B{Y}}[\ell]=\check{\B{Y}}^{1/2}[\ell]\check{\B{Y}}^{1/2}[\ell]$ where the matrix $\check{\B{Y}}^{1/2}[\ell]$ can be rewritten as  
\begin{equation*}
	\check{\B{Y}}^{1/2}[\ell]=\frac{1}{\sqrt{M}}\big((\B{\tilde{A}}\odot\B{\tilde{\Upsilon}}[\ell])\B{\tilde{D}}[\ell](\B{\tilde{A}}\odot\B{\tilde{\Upsilon}}[\ell])^H + \sigma_{\B{v}[\ell]}^2\mathbf{I}_M\big).
\end{equation*}
Since the result of applying spatial smoothing is a linear combination of the steering vectors, it is possible to identify the angles by using algorithms like \ac{MUSIC}. After computing the eigendecomposition of $\sqrt{M}\check{\B{Y}}^{1/2}[\ell]=\B{U}_k[\ell]\B{\Lambda}_k[\ell]\B{U}_k^H[\ell]$, the $L$ eigenvectors corresponding to the $L$ largest eigenvalues are discarded. Thus, we obtain an incomplete basis $\bar{\B{U}}_k[\ell]$ spanning the noise subspace and the estimator function as
\begin{equation}
\mathcal{J}_i=\sum_{\ell=1}^{N_c}\frac{1}{||\B{a}^H(\theta_{i})\B{\Gamma}^H(\theta_i)[\ell]\bar{\B{U}}[\ell]||_2^2},
\end{equation}
where $\theta_{i}$ is the $i$-th angle contained in the dictionary matrix $\B{\tilde{A}}$. Following the procedure explained for the \ac{MUSIC} algorithm, we estimate the gain variances using 
\begin{equation*}
\B{\hat{D}}[\ell]=(\B{\tilde{A}}_\mathcal{S}\odot\B{\tilde{\Upsilon}}_\mathcal{S}[\ell])^{\dag}(\check{\B{Y}}^{1/2}[\ell]-\sigma_{\B{v}[\ell]}^2\mathbf{I}_M)\big((\B{\tilde{A}}_\mathcal{S}\odot\B{\tilde{\Upsilon}}_\mathcal{S}[\ell])^{\dag}\big)^H,
\end{equation*}
or the corresponding version of \eqref{eq:gainEstimates} if the pseudo-inverse is numerically unstable.

A direct improvement of this method consists of employing all the differences in the vector $\B{y}[\ell]$ instead of discarding the repeated ones.  Notice that in \eqref{eq:vectorized_cov} the vector $\B{y}[\ell]=\operatorname{vec}(\hat{\B{C}}_{\B{\varphi}[\ell]})$ has $T_{\text{tr}}^2$ elements whereas in \eqref{eq:smoothing} only $2M-1$ elements are considered for $\check{\B{y}}[\ell]$.
Let us introduce $y_i^j[\ell]$ as the element of $\B{y}[\ell]$ denoting the $j$-th snapshot of the  $i$-th difference. In this case, we can redefine the vector $[\check{\B{y}}]_i[\ell] = \frac{1}{J}\sum_{j=1}^{J} y_i^j[\ell]$, where $J$ represents the number of snapshots available for the $i$-th difference. Note that $J$ depends on  the sparse ruler structure.

\subsection{Dictionary size and number of paths estimation}
\label{sec:DicSizeLEstim}
In this section we analyze impact of the dictionary size $G$ and the scenario where $L$ is not known in advance.

To determine a sufficient condition for a unique angular identification, we first define the Kruskal rank of a matrix $\B{Z}$, $\operatorname{krank}(\B{Z})$. If $\operatorname{krank}(\B{Z})=z$, then $z$ is the maximum number such that any $z$ columns of $\B{Z}$ are linearly independent.    
\begin{theorem}
	\label{th:musicConds}
	In the noiseless scenario with the \ac{AoD} lying on the dictionary $\B{\tilde{A}}$, and $\operatorname{rank}(\hat{\B{C}}_{\B{\varphi}[\ell]})=L$, the inequality
	\begin{align}
	L<\operatorname{krank}(\B{\Psi}[\ell]),
	\end{align}
	suffices to guarantee the unique identification of the $L$ \ac{AoD} \cite{Lee12}. See \cite{WaZi89} for the proof.
\end{theorem}

\begin{figure}
	\includegraphics[width=\columnwidth]{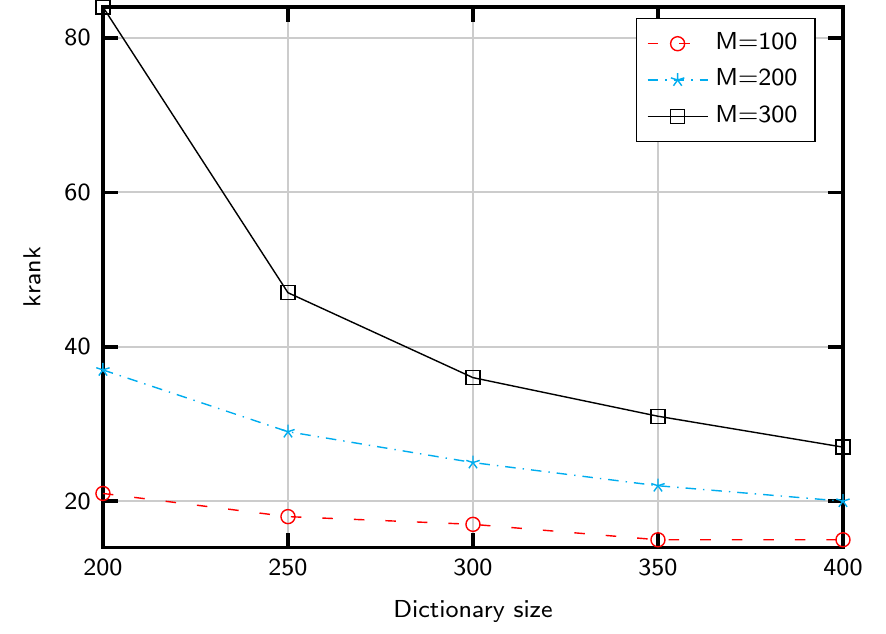}
	\caption{Kruskal rank upper bound for different array and dictionary sizes.}
	\label{fig:krank}
\end{figure}

When the training sequence is properly chosen, we can focus on the Kruskal rank of the dictionary matrix $\B{\tilde{A}}$. If $\operatorname{krank}(\B{\tilde{A}})$ is small, the probability of identifying false directions increases. Two parameters with a strong impact on the Kruskal rank are the dictionary size $G$ and the potential \acp{AoD}. To illustrate the dependence with $G$, we compute in Fig. \ref{fig:krank} an upper bound of the Kruskal rank for different dictionary and antenna array sizes. This upper bound takes into account, not any $z$ columns but only $z$ consecutive columns. This computationally feasible approximation is reasonable due to the structure of the dictionary matrices. As shown in the figure, the use of large dictionaries makes more difficult the \ac{AoD} detection using \ac{MUSIC}. 

Computational complexity of estimating gain variances also benefits from moderate dictionary sizes. Typical values are $G\approx2M$ \cite{GoRoGoCaHe18,HaCa18}. However, for inter-antenna spacing $d=\lambda/2$, the angular resolution of a \ac{ULA} is about $\frac{2}{M}$ radians \cite{Ri05}, leading to $G\approx\frac{\Delta\vartheta\pi}{360º}M$ for an angular range $\Delta\vartheta$. For instance, $G\approx M$ provides enough angular resolution for $\Delta\vartheta=120º$.

Another approach to overcome this limitation is using variations of \ac{MUSIC} like \ac{OSMP} \cite{TsLiWu18}. However, this solution might compromise the detection accuracy of clustered angles.  

In the previous methods, the number of paths $L$ was assumed to be known. Subspace estimation methods like the one in \cite{TsLiWu18} can be used to determine $L$ by comparing the differences between consecutive eigenvalues of the sample covariance matrix. In the wideband case, the information of all frequencies can be employed, showing excellent performance results (see Sec. \ref{sec:subspaceEstim}). Moreover, the computational cost over \ac{MUSIC} and \ac{SS} is negligible.

%

\section{Channel estimation}
As a means to estimate the channel, we first propose to employ the linear \ac{MMSE} estimator, as e.g. \cite{Adh13,Hon17}. For the $k$-th training block, $\B{h}[\ell]$ and $\B{v}[\ell]$ being zero mean jointly Gaussian random variables, we compute the estimator as \cite{Kay93}
\begin{equation}
	\hat{\B{h}}[\ell]=\hat{\B{C}}_{\tilde{\B{h}}[\ell]}\B{X}_k^H\left(\B{X}_k\hat{\B{C}}_{\tilde{\B{h}}[\ell]}\B{X}_k^H+\sigma_{\B{v}[\ell]}^2\mathbf{I}_{T_\text{TR}}\right)^{-1}\B{\varphi}_k[\ell].
	\label{eq:mmseChEstimator}
\end{equation}
Notice that when spatial sparsity holds, it is possible to reduce the computational complexity of the previous estimator. Indeed, when $T_\text{TR}>L$ we calculate \eqref{eq:mmseChEstimator} as
\begin{align}
\hat{\B{h}}[\ell]&=\hat{\B{C}}_{\tilde{\B{h}}[\ell]}\B{X}_k^H\left(\sigma_{\B{v}[\ell]}^{-2}\mathbf{I}_{T_\text{TR}}-\sigma_{\B{v}[\ell]}^{-4}\B{\Psi}_{\mathcal{S},k}[\ell]\right.\label{eq:lowComplMmseChEstimator}\\
&\left.\times\big(\B{\hat{D}}^{-1}_{\mathcal{S}}[\ell]+\sigma_{\B{v}[\ell]}^{-2}\B{\Psi}_{\mathcal{S},k}^H[\ell]\B{\Psi}_{\mathcal{S},k}[\ell]\big)^{-1}\B{\Psi}_{\mathcal{S},k}^H[\ell]\right)\B{\varphi}_k[\ell].\notag	
\end{align}

\subsection{Delay and gain estimations}
In this section we explain how to estimate the delay matrices $\B{\Theta}[\ell]$ and the variance gain vectors $\tilde{\B{g}}$, cf. \eqref{eq:wideChAngDomain}. 
Delay estimation methods have been proposed in a different context \cite{WiKiPr98}. Since the delays are wide sense stationary, we propose to estimate them from the observations. Then, a robust and low complexity estimator is calculated for the frequency selective channels.

Recall that $\B{\varphi}_k[\ell]\approx\B{\Psi}_{\mathcal{S},k}[\ell]\B{\Theta}[\ell]\tilde{\B{g}}+\B{v}[\ell]$.  We define the vector $\B{\zeta}[\ell]=\B{\Theta}[\ell]\tilde{\B{g}}$ containing the channel gains for subcarrier $\ell$. We can again define the linear \ac{MMSE} estimator as
\begin{align}
	\label{eq:widebanGains}
	\hat{\B{\zeta}}[\ell]=\B{\zeta}[\ell]+&\breve{\B{\zeta}}[\ell]=\hat{\B{D}}_{\mathcal{S}}[\ell]\B{\Psi}_{\mathcal{S},k}^H[\ell]\\
	&\times\left(\B{\Psi}_{\mathcal{S},k}[\ell]\hat{\B{D}}_{\mathcal{S}}[\ell]\B{\Psi}^H_{\mathcal{S},k}[\ell]+\sigma_{\B{v}[\ell]}^2\mathbf{I}_{T_\text{TR}}\right)^{-1}\B{\varphi}_k[\ell],\notag
\end{align}
where $\breve{\B{\zeta}}[\ell]$ is the estimation error.
We now collect the estimated gains for all the frequencies associated to the  $l$-th path in a vector and multiply it times the first $N$ columns of the \ac{DFT} matrix $\B{F}\in\mathbb{C}^{N_c\times N}$ to get
\begin{align}
	\B{\zeta}_{l}&=\B{F}^H\begin{bmatrix}[\hat{\B{\zeta}}[1]]_l,\ldots,[\hat{\B{\zeta}}[N_c]]_l\end{bmatrix}^T\notag\\
	&=\tilde{g}_{l}\B{p}_{\text{rc}}(\tau_l)+\B{w},
\end{align} 
 which contains $N$ noisy samples of the function $p_{\text{rc}}(\tau_l)$, with  $\B{w}\sim\mathcal{N}_\mathbb{C}(\mathbf{0},\sigma^2_{\B{w}}\mathbf{I}_{N_c})$ the estimation noise assuming that the error for different subcarriers is statistically independent, and the pulse samples stacked in 
 $\B{p}_{\text{rc}}(\tau_l)=[p_{\text{rc}}(-\tau_l),p_{\text{rc}}(T_s-\tau_l),\ldots,p_{\text{rc}}\big((N-1)T_s-\tau_l\big)]^T$. 
These samples are, however, scaled by the unknown complex channel gain $\tilde{g}_{l}$.

We now denote the $k$-th observation of $\B{\zeta}_{l}$ as $\B{z}_{l,k}$. Hence, since $p_{\text{rc}}(\cdot)$ is known, we build the estimator function
\begin{align}
	\mathcal{D}(\tau)&=\sum_{k=1}^K\frac{|\B{p}_{\text{rc}}^T(\tau)\B{z}_{l,k}|}{\|\B{p}_{\text{rc}}(\tau)\|_2\|\B{z}_{l,k}\|_2}.
	\label{eq:tauEstimator}
\end{align}
Unfortunately, $\mathcal{D}(\tau)$ is a not monotonic function. Thus, we minimize $\mathcal{D}(\tau)$ over $\tau$, e.g. by performing a linear search over $[0,(N-1)T_s]$ to obtain the delay estimate $\hat{\tau}_{l}$. In the high SNR regime, it is enough to sound within the interval corresponding to the two samples with larger absolute values. 

Once provided the delays for each path $\hat{\tau}_l$, we compute $\hat{\B{\Theta}}[\ell]$ for all frequencies. These matrices are next used to estimate the frequency flat channel gains $\tilde{\B{g}}_k$, associated to each propagation path. Introducing  $\B{z}_k[\ell]$ as the $k$-th observation of $\hat{\B{\zeta}}[\ell]$ in \eqref{eq:widebanGains}, we build the gain estimator as follows
\begin{align}
\hat{\B{g}}_k = 
\begin{bmatrix}
\hat{\B{\Theta}}[1]\\
\vdots\\
\hat{\B{\Theta}}[N_c]
\end{bmatrix}^\dag
\begin{bmatrix}
\B{z}_k[1]\\
\vdots\\
\B{z}_k[N_c]
\end{bmatrix},
\label{eq:gainsEstimator}
\end{align}
which is the minimum variance unbiased estimator \cite{Kay93}.

Employing both the matrices $\hat{\B{\Theta}}[\ell]$ and the gains $\hat{\B{g}}_k$, we alternatively calculate an estimate of the channel as
\begin{equation}
	\hat{\B{h}}[\ell]=(\B{\tilde{A}}_{\mathcal{S}}\odot \B{\tilde{\Upsilon}}_{\mathcal{S}}[\ell])\hat{\B{\Theta}}[\ell]\hat{\B{g}}_k.
	\label{eq:channelEstimationDelay}
\end{equation}
This method employs the observations for all the subcarriers to estimate the gains of each propagation path. Thus, even in the case where certain frequencies present very poor SNR, their corresponding channel can be estimated by using the information from the remaining frequencies. 

\section{Simulation Results}

The following setup is considered for the numerical experiments. We assume constant training, i.e. $\B{X}_k = \B{X}, \forall k$ generated from a Wichmann ruler of length $T_{\text{tr}}$. The number of transmit antennas is $M=200$, and the dictionary size is $G=400$ with equally spaced angles within the range $[\pi/3,2\pi/3]$. We consider the channel covariance model of  \eqref{eq:vchannel_model} with $L$ propagation paths uniformly distributed in $[\pi/3,2\pi/3]$; different number of snapshots $K=\{5,10,20,50,100,150,200\}$, and $N=8$ delay taps uniformly distributed in $[0,N-1]T_s$, with $T_s=1/f_s$ and $f_s=1760$ MHz. The numerical results are averaged over $200$ channel realizations. To evaluate the accuracy of the covariance identification, we employ the \ac{NMSE} metric, defined as
\begin{equation}
\text{NMSE}=\frac{\|\hat{\B{C}}_{\B{h}}-\B{C}_{\B{h}}\|_F^2}{\|\B{C}_{\B{h}}\|_F^2},\notag
\end{equation}
with the Frobenius norm $\|\cdot\|_F$. 
In the case of channel estimation, we propose to use the efficiency $\eta\in[0,1]$, similar to the figures of merit in e.g. \cite{HaCa17,PaHe17,ZhZhSa18}. This insightful metric evaluates the signal power lost by performing beamforming using the estimated channels, and it is given by
\begin{equation}
\eta=\frac{|\hat{\B{h}}^H\B{h}|}{\|\hat{\B{h}}\|\|\B{h}\|}.
\end{equation} 
In the wideband scenario, the metrics are averaged over the number of subcarriers $N_c$.
\subsection{Narrowband Covariance Identification}

\begin{figure}
\includegraphics[]{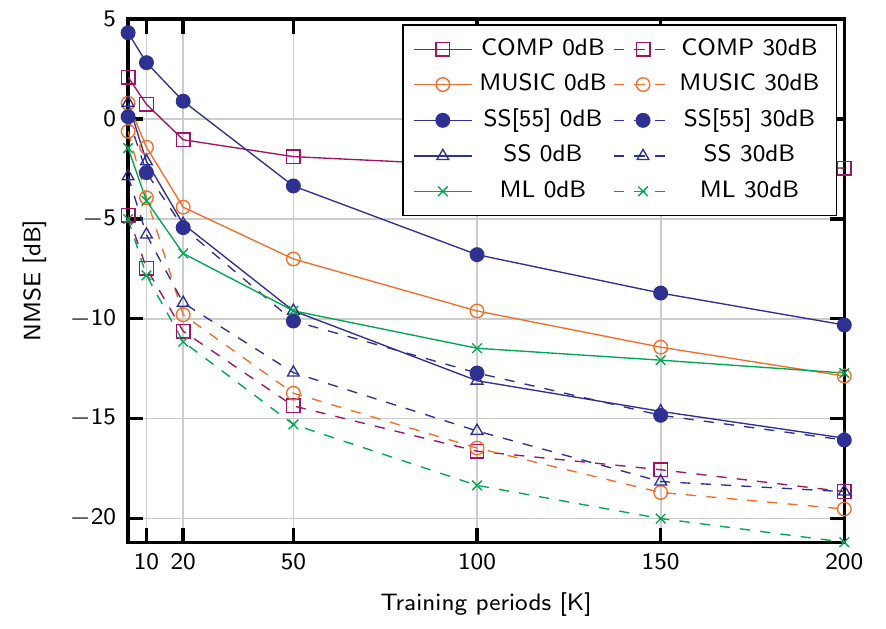}
\caption{NMSE of different covariance identification strategies for $M=200$ antennas in the narrowband case, and $L=15$ propagation paths.}
\label{fig:FigNMSE_Narrow_0-30dB}
\end{figure}

Fig. \ref{fig:FigNMSE_Narrow_0-30dB} shows the \ac{NMSE} for the channel covariance matrices determined using the algorithms described in the previous section for SNRs of $0$ dB and $30$ dB. In this experiment, we compare the proposed methods with the benchmark \ac{ML} in the narrowband scenario. Also, the performance gain of \ac{SS} with respect to \cite{ArLe13} is shown. It is remarkable the robustness of \ac{MUSIC} compared to \ac{COMP} in the low SNR regime. \ac{ML} performance in terms of \ac{NMSE} is excellent, but it is also the most computationally expensive. Moreover, \ac{MUSIC} and \ac{SS} perform better at $0$ dB for large number of snapshots $K$.

\subsection{Wideband Covariance Identification}

\begin{figure}
	\includegraphics[]{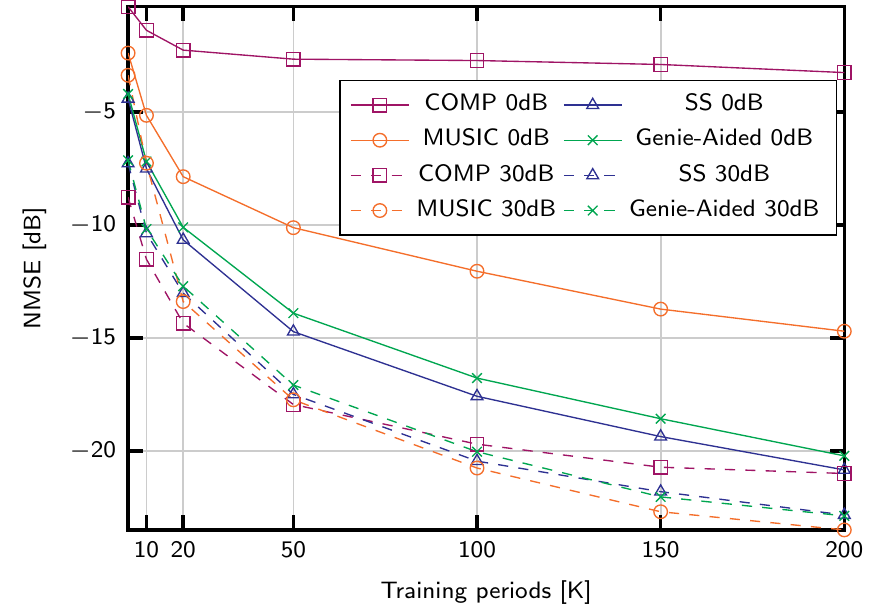}
	\caption{NMSE of different covariance identification strategies for $M=200$ antennas in the wideband case, and $L=15$ propagation paths.}
	\label{fig:FigNMSE_Wide_Sparsity15}
\end{figure}

\begin{figure}
\includegraphics[]{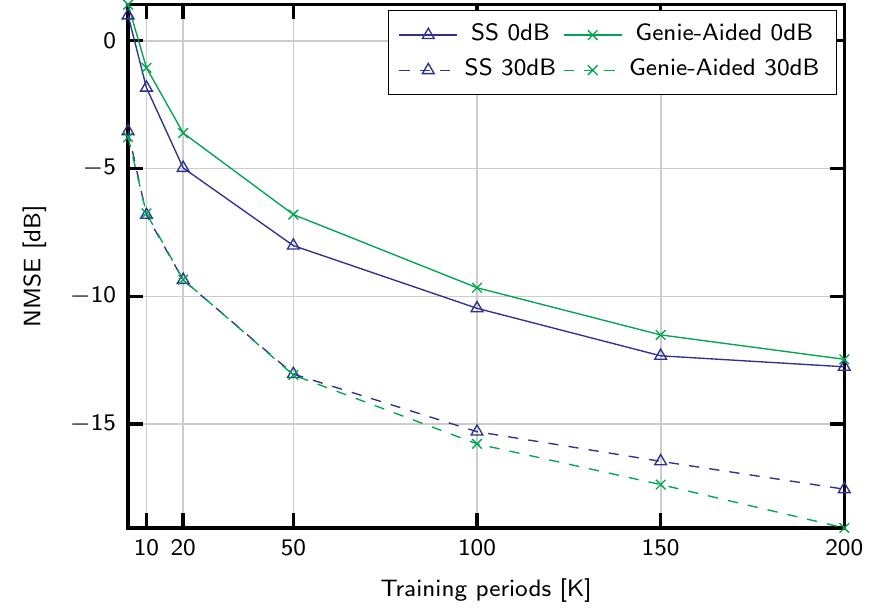}
\caption{NMSE of different covariance identification strategies for $M=200$ antennas in the wideband case, and $L=70$ propagation paths.}
\label{fig:FigNMSE_0dB_Sparsity70}
\end{figure}

Figure \ref{fig:FigNMSE_Wide_Sparsity15} shows the NMSE obtained by some of the algorithms described for the wideband case with a \ac{ULA} channel model. The training  length is $T_{\text{tr}}=50$. Results for \ac{SS} are obtained with the improvement explained at the ending of Subsection \ref{subsec:smooth}. Both \ac{SS} and \ac{MUSIC} improve the estimation accuracy by exploiting the symmetry of the wideband gain covariances. As benchmark, we include the Genie-Aided curve that individually estimates the gain covariances for each subcarrier using \eqref{eq:gainEstimates} for known channel angles.  

Figure \ref{fig:FigNMSE_0dB_Sparsity70} shows the NMSE for sparsity $L=70$ for two levels of SNR. \ac{MUSIC} and \ac{DCOMP} are not suited for this scenario because of the limited rank of the sample covariance matrices, and due to the lack of angular sparsity. The performance of \ac{SS} is better than the one obtained with Genie-Aided strategy for low SNRs, but performs slightly worse for high SNR and long training sequences.

\subsection{Channel Subspace Estimation}
\label{sec:subspaceEstim}
\begin{figure}
	\includegraphics[]{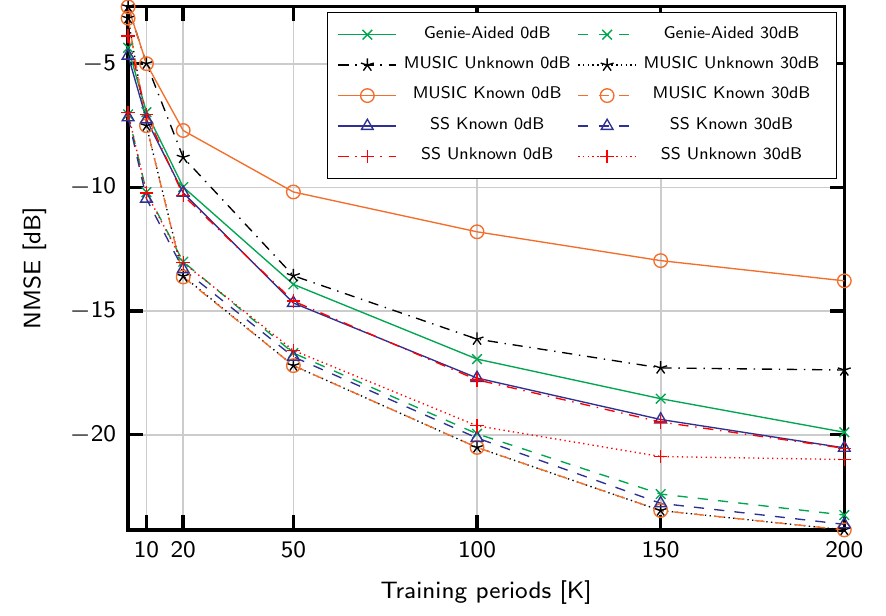}
	\caption{NMSE of different channel estimation strategies for $M=200$ antennas, $L=15$ propagation paths, and subspace estimation.}
	\label{fig:FigNMSEChannel_Wide_Sparsity_Unkown}
\end{figure}
Figure \ref{fig:FigNMSEChannel_Wide_Sparsity_Unkown} shows the robustness against uncertainty in the number of channel propagation paths. With regard to estimating the number of paths, we employ Alg. 2 in \cite{TsLiWu18}, empirically adjusting the parameter for different identification methods and SNR regimes. The performance obtained with subspace estimation is similar to that of the strategies with known $L$. Remarkably, including more angles than the actual number of channel paths improves the performance of \ac{MUSIC} for $0$ dB of SNR because some of the angles are incorrectly discarded. Finally, the performance of \ac{SS} is slightly worse for $30$ dB and long training periods.

\subsection{Wideband Channel Estimation}
\begin{figure}
\includegraphics[]{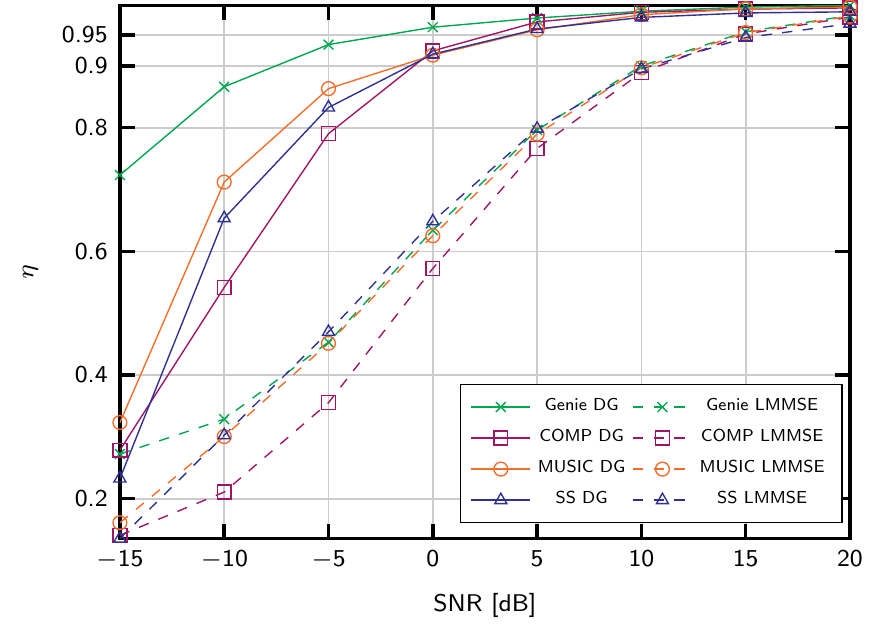}
\caption{Efficiency of  channel estimation for $M=200$ antennas, $L=15$ propagation paths, $T_\text{TR}=25$ training length, and $K=100$ snapshots.}
\label{fig:FigNMSEChannel_Wide_Sparsity15}
\end{figure}

\begin{figure}
\includegraphics[]{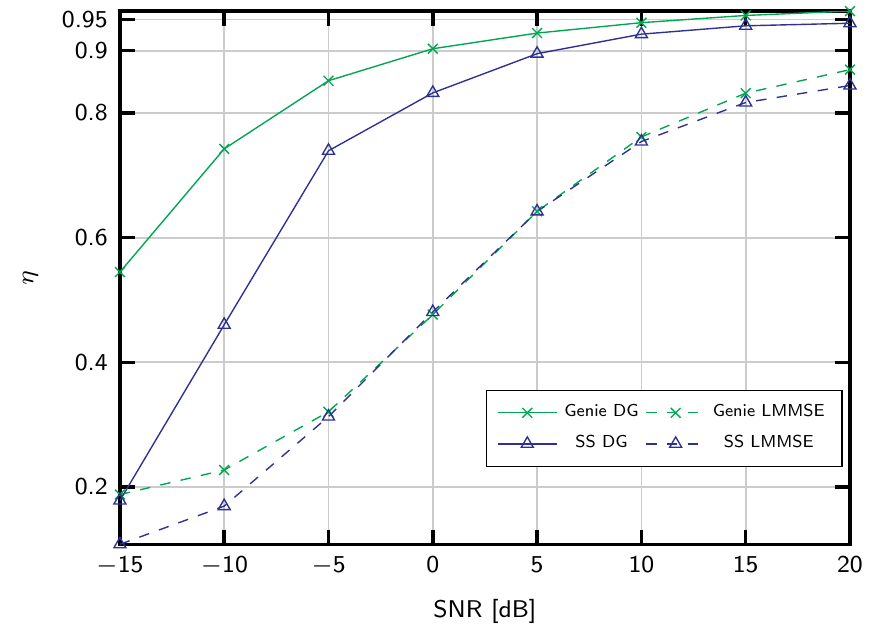}
\caption{Efficiency of channel estimation for $M=200$ antennas,  $L=15$ propagation paths, $T_\text{TR}=25$ training length, and $K=100$ snapshots.}
\label{fig:FigNMSEChannel_Wide_Sparsity35}
\end{figure}

Figures \ref{fig:FigNMSEChannel_Wide_Sparsity15} and \ref{fig:FigNMSEChannel_Wide_Sparsity35} show the performance of the channel estimator with the covariance identified by different methods. We consider two channel estimation methods: the conventional LMMSE, given by \eqref{eq:mmseChEstimator}, and the one based on the estimation of the delay and the gains, cf. \eqref{eq:channelEstimationDelay}, which is denoted as DG. The number of channel blocks is set to $K=100$ and the number of delay candidates $\tau$ considered in \eqref{eq:tauEstimator} is $20{N}$. Since the proposed DG method uses the information of all the subcarriers to produce the estimates, it is very robust against noise. The difficult scenario where the rank of the sample covariance matrix is smaller than the number of paths is shown in Fig.  \ref{fig:FigNMSEChannel_Wide_Sparsity35}. Again, the proposed DG strategy provides more accurate estimates than LMMSE. Indeed, more than $75\%$ of the signal power is captured at $-5$dB with \ac{SS}.

\subsection{Computational complexity}
\label{sec:complexity}
 
\setlength{\tabcolsep}{0.3em}
\renewcommand*{\arraystretch}{1.5}
\begin{table}[h]
	\caption{Complexities of covariance identification methods }
	\vspace*{-0.1cm}
	\label{tab: covComplexities}
	\begin{footnotesize}
		\begin{center}
			\begin{tabular}{|c|c|}
				\cline{2-2}
				\multicolumn{1}{c|}{} & {\textbf{Covariance Identification}} \\\cline{2-2}\hline
				MUSIC(a) & $\mathcal{O}\big(N_c(T_\text{Tr}^3+(T_\text{Tr}^2+T_\text{Tr})G+L^2T_\text{Tr}+2LT_\text{Tr}^2)\big)$ \\\hline
				MUSIC(b) & $\mathcal{O}\big(N_c(T_\text{Tr}^3+(T_\text{Tr}^2+T_\text{Tr})G+L^4T_\text{Tr}^2+L^2T_\text{Tr}^2)\big)$ \\\hline
				DCOMP  & $\mathcal{O}(N_c(2KT_\text{Tr}^2G+KL^3T_\text{Tr}+2KT_\text{Tr}^2(L^2+L)+KLT_\text{Tr}^3)\big)$ \\\hline
				SS(a) & $\mathcal{O}\big(N_c(M^2+M^3+(M^2+M)G+L^2M+2LM^2)\big)$  \\\hline
				SS(b) & $\mathcal{O}\big(N_c(M^2+M^3+(M^2+M)G+L^4M^2+L^2M^2)\big)$  \\\hline
				\multicolumn{1}{c|}{} & {\textbf{Computing  Estimator}} \\\hline
				DG & $\mathcal{O}\big(N_c(3L^2T_\text{Tr}+T_\text{Tr}^3+NL)+KNT\big)$\\\hline
				Eq. \eqref{eq:mmseChEstimator} & $\mathcal{O}(N_c(3M^2T_\text{Tr}+T_\text{Tr}^3)\big)$\\\hline
				Eq. \eqref{eq:lowComplMmseChEstimator} & $\mathcal{O}\big(N_c(M^2T_\text{Tr}+3T_\text{Tr}L^2+L^3)\big)$\\\hline
				\multicolumn{1}{c|}{} & {\textbf{Channel Estimation}} \\\cline{2-2}\hline
				DG & $\mathcal{O}\big(N_c(LT_\text{Tr}+KL^2+ ML+L^2)\big)$\\\hline
				LMMSE & $\mathcal{O}(N_cMT_\text{Tr})$\\\hline				
			\end{tabular}
		\end{center}
	\end{footnotesize}
\end{table}

We summarize the computational complexity of the proposed methods in Table \ref{tab: covComplexities}. Recall that a reasonably fast covariance identification is important, but not as critical as the actual channel estimation. Indeed,  we only employ a training stage contrarily to other approaches like \cite{Adh13,HuHuXuYazeroNorm}. Therein, the estimated covariance is employed to design an additional training sequence and finally estimate the channel.

\ac{MUSIC}(a) and SS(a) are for the favorable scenario where the gain variances are estimated using well conditioned measurement matrices, whilst \eqref{eq:gainEstimates} with the Khatri–Rao product is considered for \ac{MUSIC}(b) and SS(b).
The complexities have been calculated assuming $T_\text{Tr}>L$ for \ac{MUSIC} and \ac{DCOMP}. It is noticeable that neither \ac{MUSIC} nor \ac{DCOMP} depend on the number of antennas, although both of them linearly depend on the dictionary size $G$. The cyclic prefix length $N$ is also relatively small, and $T$ is the number of $\tau$ candidates evaluated to estimate the delays using \eqref{eq:tauEstimator}. Since \ac{DCOMP} and \ac{MUSIC} apply to scenarios where $L$ and $T_\text{Tr}$ are very small compared to $M$ or $G$, the complexity of the latter is smaller in general. Further, the number of steps performed by \ac{DCOMP} is larger and  its computational burden depends on the number of snapshots $K$. 

When $L\centernot\ll M$, the computational complexity virtually depends on the number of antennas. Hence, \ac{SS} and \ac{MUSIC} establish a tradeoff among $T_\text{TR}$, $L$ and $M$, since we could use the less complex option by adjusting $T_\text{TR}$ and $K$.

We now differentiate between the operations computed once using the estimated covariance matrix, referred to as ``Computing Estimator'', and the channel estimation for each observation, referred to as ``Channel Estimation''. Remarkably, the complexity of the proposed estimator (DG) does not depend on $M$, cf. \eqref{eq:widebanGains}-\eqref{eq:tauEstimator}. Contrarily, the LMMSE estimator quadratically depends on $M$ for \eqref{eq:mmseChEstimator} or \eqref{eq:lowComplMmseChEstimator}. If we consider time-variant training, these complexities scale by $K$. 

Regarding the actual channel estimation, in scenarios suitable for \ac{MUSIC} or \ac{DCOMP}  with $L<T_\text{Tr}$, DG channel estimation is more efficient. This applies to \ac{mmWave} frequencies and other massive \ac{MIMO} scenarios,  e.g. \cite{GaDaWaCh15,ShZhAlLe16,HuHuXuYa17,ZhWaSu17,RaLa14,ChSuLi17}. Contrarily, conventional LMMSE estimator might be cheaper for large number of channel propagation paths, i.e. $L>\sqrt{M}$.

\section{Conclusions}
In this work we propose a channel estimation method based on the covariance identification in \ac{FDD} wideband massive \ac{MIMO} systems. The covariance identification method exhibits a good performance in terms of \ac{NMSE} compared to other methods in the literature, even for very short training sequences. Moreover, the channel estimator developed in this work is very robust against noise and its computational complexity is independent of the number of antennas. These facts are key to reduce the channel training overhead. Finally, the performance of the designed methods is illustrated for a broad variety of scenarios.

\bibliographystyle{IEEEtranTCOM}
\bibliography{bibliography}

\end{document}